\documentclass{bmvc2k}
\usepackage{amsmath, amsthm, amssymb, amsfonts, amsbsy,amssymb,fixmath}
\usepackage{epstopdf}
\usepackage{upgreek}
\usepackage{booktabs}
\usepackage{multirow}
\usepackage{array}
\usepackage{float}
\usepackage{kantlipsum}
\usepackage{pifont}
\usepackage{graphicx}
\usepackage{caption}
\usepackage{hyperref}
\usepackage{algorithm}
\usepackage{algpseudocode}
\usepackage{tikz}
\newcommand{\norm}[1]{\left\lVert#1\right\rVert}
\def\cbr#1{\left\lbrace #1 \right\rbrace} 
\def\sbr#1{\left[ #1\right]} 
\def\nbr#1{\left( #1\right)}

\def\bf #1{\mathbf{#1}}
\def \p #1{\mathbf{#1}_i}
\def \B #1{\textbf{#1}}

\title{Patch-based Interferometric Phase Estimation via Mixture of Gaussian Density Modelling $\&$ Non-local Averaging in the Complex Domain}

\addauthor{Joshin P. Krishnan}{joshin.krishnan@lx.it.pt}{1}
\addauthor{Jos\'e M. Bioucas-Dias}{bioucas@lx.it.pt}{1}

\addinstitution{
Instituto de Telecomunica\c c\~oes\\
Instituto Superior T\'ecnico\\
Universidade de Lisboa\\ 
Portugal
}

\runninghead{Joshin P. K., J. Bioucas-Dias}{Interferometric Phase Estimation}

\def\eg{\emph{e.g}\bmvaOneDot}

\begin{document}
\maketitle

\begin{abstract}
This paper addresses interferometric phase (InPhase) image denoising, i.e., the denoising of phase modulo-2$\pi$ images from sinusoidal 2$\pi$-periodic and noisy observations. The wrapping discontinuities present in the InPhase images, which are to be preserved carefully, make InPhase denoising a challenging inverse problem. We propose a novel two-step algorithm to tackle this problem by exploiting the non-local self-similarity of the InPhase images. In the first step, the patches of the phase images are modelled using Mixture of Gaussian (MoG) densities in the complex domain. An Expectation Maximization (EM) algorithm is formulated to learn the parameters of the MoG from the noisy data. The learned MoG is used as a prior for estimating the InPhase images from the noisy images using Minimum Mean Square Error (MMSE) estimation. In the second step, an additional exploitation of non-local self-similarity is done by performing a type of non-local mean filtering. Experiments conducted on simulated and real (MRI and InSAR) data sets show results which are competitive with the state-of-the-art techniques.
\end{abstract}

\section{Introduction}
\label{sec:intro}

Phase imaging systems play a vital role in many present day technologies, namely in the field of surveillance, remote sensing, medical diagnostic, weather forecasting and photography. Often, in such systems, a physical quantity of interest is coded in an image of phase using a suitable coherent imaging techniques.

Popular and relavant technologies in this categorie include Interferometric Synthetic Aperture Radar \& Sonar (InSAR/InSAS) \citep{1974_Graham_Synthetic}, \cite{1986_Zebker_Topographic}, \citep{1998_Ghiglia_Two}, \cite{2000_Rosen_Synthetic}, \cite{2002_Dias_Z}, Magnetic Resonance Imaging (MRI) \cite{1973_Lauterbur_Image}, \cite{1992_Hedley_new}, Optical Interferometry  \cite{1994_SPandit_Data} and High Dynamic Range (HDR) Photography \cite{1994_SPandit_Data}. For \eg, in InSAR/InSAS, the radar/sonar signals scattered from a terrain are collected using spatially  distant sensors. The information related to the topography of the terrain is coded in the phase differences of the signals collected at the different sensors. Infact, the spatial diversity of the paths of the received signals are exploited and thereby the terrain topography is decoded \cite{1997_Griffiths_Interferometric}. In MRI, phase estimation is required to measure the magnetic field deviation
maps, which can be used to correct echo-planar image geometric
distortions \cite{1995_Jezzard_Correction}, to determine chemical shift based thermometry \cite{2000_Quesson_Magnetic}, and to implement BOLD contrast based venography \cite{2003_Rauscher_Automated}. In optical interferometry, the shape, deformation, and vibration of the objects are measured using phase estimation \cite{1994_SPandit_Data}.    
Phase unwrapping algorithms are used in HDR photography to recover very high range radiance levels from a single modulus image of limited bit depth.

Since the phase is closely linked with the wave propagation phenomenon, the measured signals depend only on the principal (wrapped) values of the original phase (absolute phase), which we term as interferometric phase, usually defined in the interval $\left[ -\pi, \pi \right)$. The interferometric phase is thus a sinusoidal and non-linear function of the absolute phase, which renders absolute phase estimation a hard inverse problem.  In addition, the interferometric phase is usually corrupted by the noise introduced by the acquisition mechanism and electronic equipments, which further complicates the inverse problem which is the inference of the absolute phase from interferometric measurements.
This problem is often tackled in a two-step approach. In the first step, denoising of the noisy wrapped phase is taken care and in the second step, the denoised phase image is unwrapped. InPhase image denoising should be addressed with special care since the wrapping discontinuities should be preserved carefully for the second stage of unwrapping.

The local polynomial approximation (LPA) to InPhase image denoising consists in assuming that the absolute phase is well approximated by a low order polynomial in small windows \citep{2006_katkovnik_local}. Though LPA performs well in areas of smooth phase variation, it results in over-smoothing in those areas where the phase variation is large or there are discontinuities. Another conventional but promising tool for wrapped phase denoising is the time-frequency analysis \cite{2007_Kemao_Twodimensional}, \citep{2007_Kemao_Comparative} based filtering. Here the windowed Fourier transform (WFT) of the phase surface is considered and exploits the fact that quite often the WFT of the complex phase is clustered in a small set of frequencies, i.e., the WFT coefficients are well approximated by sparse representations.

In both of the aforementiond techniques, i.e. LPA \citep{2006_katkovnik_local} and WFT \cite{2007_Kemao_Twodimensional}, \citep{2007_Kemao_Comparative} based, the size of the window plays a key role. An oversized window damages the essential patterns of the phase image whereas a very small window fails to perform effective denoising action. One way to address this issue by adapting the size of the local windows based on phase smoothness and the noise level. PEARLS algorithm \citep{2008_Bioucas_Absolute} successfully addresses this issue by incorporating a first order LPA using adaptive window size \citep{2006_katkovnik_local}. But the first order polynomial limits the denoising performance in phase surfaces containing discontinuities. 

The NL-InSAR method introduced in \citep{2011_Deledalle_NL} is state-of-the-art. This method exploits the non-local self-similarity existing in most real world images. A well known algorithm exploiting this property of the images of the real world is the block matching with 3D filtering (BM3D) \citep{2007_Dabov_Image}, in which similar patches of the images are grouped together and collaborative filtering is applied. Another recent approach that exploits the non-local self similarity consists in learning a dictionary in which patches are well approximated by linear combinations of a few atoms taken from that dictionary\citep{2006_elad_image}. Sparse regression on dictionaries identifies a low dimensional sub-spaces of clean patches and implicitly projects the noise from a high to a low dimensional subspace \citep{2015_Hongxing_Interferometric}. This gives rise to a large noise reduction since, in the case of independent and identically distributed (iid) noise, the power of the projected noise is proportional to the dimension of the subspace.
 
In this paper, we propose a novel approach to address InPhase image denoising by exploiting the non-local self-similarity of the phase images in a two-step algorithm. In the first step, the patches of complex phase images are modelled using MoG densities in the complex domain. Due to the non-local self-similarity of the complex phase images, the clean patches are well modelled by very few eigen-directions of the covariance matrices of the MoG components. In other words, the first step exploits the eigenspace based sparsity of the phase patches. The parameters, i.e., the covariance matrix, mean and mixing coefficients of the MoG are learned from complex domain patches of the noisy data. The learned MoG is then used as a prior for estimating the interferometric phase images from the noisy ones. The main contribution of the first step of our work, which is inspired from the recent state-of-the-art image denoising techniques based on MoGs (see, e.g. \cite{2015_Teodoro},\cite{2013_wang_sure}), can be summarized as follows: 1) an algorithm to learn the probabilistic model; this is accomplished by designing an Expectation Maximization (EM) algorithm for MoG densities in the complex domain;
 2) a Minimum Mean Square Error (MMSE) based estimation technique; this is to estimates the clean patches from the noisy ones using the learned model. 

The second step further exploits the non-local self-similarity; for each patch estimated in MoGInPhase, a weighted patch average is carried out in which more similar patches are given higher weights. To accomplish this, we use $\textit{l}_2$ distance between patches as a measure of similarity for designing the weights. We term the first step as MoGInPhase and the second step as Non-local averaging (NL-averaging).
\begin{figure}[h]
\begin{center}
\includegraphics[height=4cm,width=13cm]{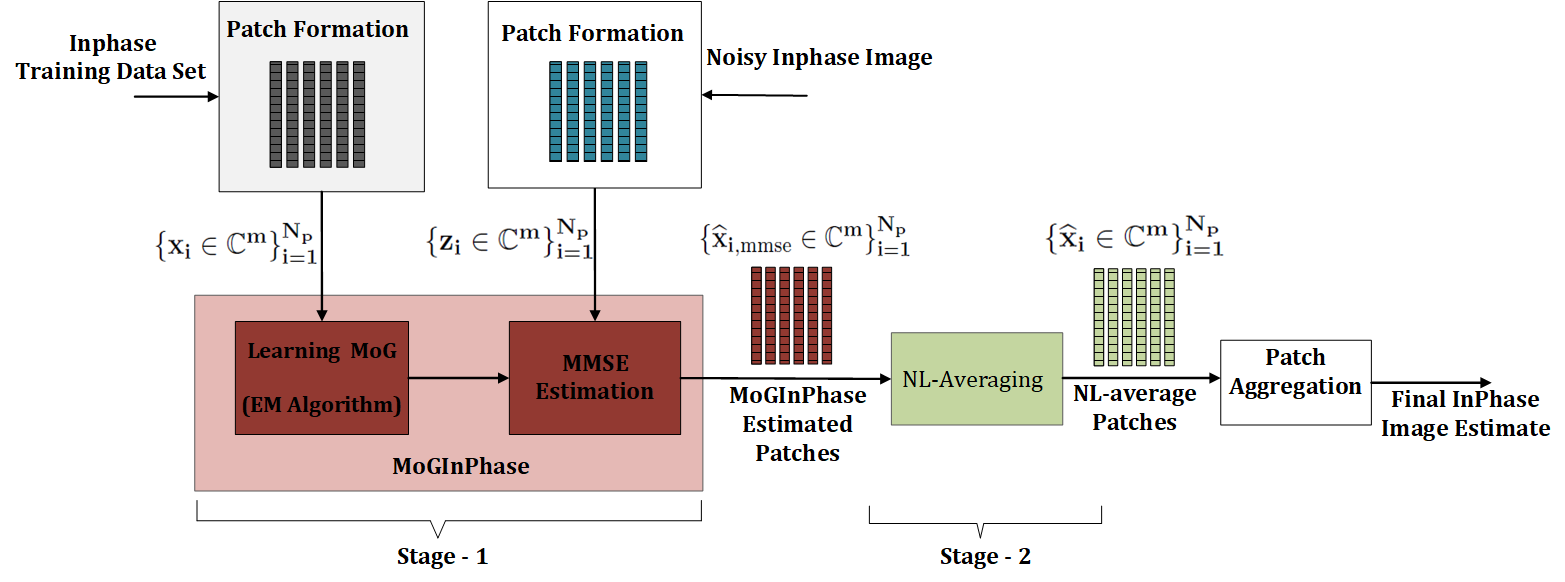}
\label{surface}
\end{center}
\caption{Block diagram representation of the proposed InPhase denoising algorithm}
\end{figure}
\vspace{-1.0cm}
\section{Problem Formulation}
\label{sec:format}
Herein, we assume the following observation model at a given image pixel:
\begin{equation}
\label{eqn:1}
z=ae^{j\upphi}+n,\ j=\sqrt{-1},
\end{equation}
where $a\geq0$, $\upphi$, and $n = n_{I} + jn_{Q}$ are the values of the amplitude, phase, and  complex domain noise of the image at the given pixel. The noise is assumed to be zero-mean Gaussian circular  (details on circularity is given later) and white  with variance $\sigma^2$.
Although the observation model may vary among the different phase imaging modalities, model \eqref{eqn:1} captures the essence of the problem. We assume that the phase image is defined on a grid of size $N := N_1 \times N_2$. The observed noisy pixels are arranged into a column vector $\bf{z} := [z_i, i = 1, . . . ,N]^T$ according to the lexicographical order of the set $\lbrace1, . . . ,N_1\rbrace \times \lbrace1, . . . ,N_2\rbrace$. In a similar way, we form the column vector for the true amplitude image ($\bf a \in \mathbb{R}^N $), true phase image ($\boldsymbol \upphi \in \mathbb{R}^N $) and noise image ($\bf n \in \mathbb{C}^N $), i.e., $\bf a := [a_i, i = 1, . . . ,N]^T$, $ \boldsymbol{\upphi} := [\upphi_i, i = 1, . . . ,N]^T$ and $\bf n := [n_i, i = 1, . . . ,N]^T$. The interferometric phase $\boldsymbol{\upphi}_{2\pi}=[\upphi_{2\pi_i}, i = 1, . . . ,N]^T$ is defined as 
\begin{eqnarray}
\upphi_{2\pi}:=\mathcal{W}( \boldsymbol{\upphi}),
\end{eqnarray}
where $\mathcal{W}(.)$ is the wrapping operator that performs component wise 2$\pi$-modulo wrapping operation defined by
\begin{eqnarray}
\mathcal{W}:\mathbb{R} &\longrightarrow & \left[ -\pi, \pi\right) \nonumber \\ 
\boldsymbol \upphi & \longrightarrow & \mod(\boldsymbol  \upphi+\pi, 2\pi)-\pi, \label{eqn:wrap}
\end{eqnarray}
where the function$\mod(. , 2\pi)$ denotes modulo $2\pi$ function. Let a clean complex patch be denoted as $\bf x:=[x_i, i = 1, . . . ,N]^T$ with $x_i = a_i e^{j\phi_i}$. We remark that $\boldsymbol \upphi_{2\pi}$ = arg$(\bf x)$. The denoising strategy that we propose is patch-based. We adopt the the method followed in \cite{2015_Hongxing_Interferometric} for the formation and aggregation of the patches. Consider a complex noisy observation $\mathbf{z} \in \mathbb{C}^N$. All possible patches of size $\sqrt{\mathbf{m}} \times \sqrt{\mathbf{m}} $ are formed which is denoted as $ \mathbf{z}_i \in \mathbb{C}^\mathbf{m} $ where $i$ is the index corresponding to the centre pixel of the patch. The total number of overlapping patches is $N_p = (N_1 - \sqrt{m} + 1)(N_2 - \sqrt{m} + 1)$. Let $ \mathbf{x}_i \in \mathbb{C}^m $ and $ \mathbf{n}_i \in \mathbb{C}^m $ be the $i^{th}$ patches of $\mathbf{x}$ and $\mathbf{n}$ respectively. With the above notations in place, we define the InPhase estimation as the estimation of $\mathbf{x}_i$ from the noisy observation $\mathbf{z}_i$ where
\begin{equation}
\mathbf{z}_i=\mathbf{x}_i+\mathbf{n}_i,\ \ \ i=1,. . .,N_p. \label{eqn:patch}
\end{equation}
\vspace{-1.1cm}
\section{MoGInPhase Estimation}
\label{sec:MoG}
The complex vectors of patches $\mathbf{x_i} \in \mathbb{C}^m$ are modelled with a mixture of circularly symmetric Gaussian densities. An EM algorithm is formulated to learn the parameters of the MoG from the noisy samples. The learned MoG is used as a prior to compute the MMSE estimates of the clean patches from the noisy ones.
\vspace{-0.3cm}
\subsection{Circular-Symmetric Assumption}
\label{sec:CS}

Let $\bf {X}=\sbr{\bf {x}_1,..., \bf {x}_{N_p}}^T$ be complex jointly-Gaussian random vectors. The key idea of MoGInPhase is that the non-local and self-similar phase patches are modelled by few eigen-directions of the covariance matrices of the MoG components. In order to have an efficient representation using MoG, in the context of patch-based phase inference, we assume a rotationally invariant probability distribution, i.e., a patch $\bf {x}_i$ and an another patch $\bf {x}_j=e^{j\gamma }\bf {x}_i$ (with a common phase shift $\gamma$ to all of its pixel) should have the same probability distribution for any given $\gamma \in \mathbb{R}$. This motivates the assumption of circular-symmetry to the components of the MoG. The probability distribution of the complex domain phase patch $\mathbf{x}_i=\mathbf{x}_{i\Re}+j\mathbf{x}_{i\Im}$ is defined as the joint probability distribution of its real and imaginary parts\citep{2006_Gallager_Circular}, i.e., the distribution of the random variable $\tilde{\mathbf{x}_i}=[\mathbf{x}_{i\Re},\mathbf{x}_{i\Re}]^T\in  \mathbb{R}^{2 m}$.  With the assumption of circular symmetry, it is straightforward to prove that the mean $ \sbr{ \mu_{\bf {x}_i}=E\nbr{\bf {x}_i}, E \ \text{is the expectation operator}}$ and the pseudo covariance matrix $\sbr{M_{\bf {x}_i}=E\cbr{ ({\bf {x}_i}-\mu_{\bf {x}_i})({\bf {x}_i}-\mu_{\bf {x}_i})^T}}$ are zeros. This simplifies the expression of circular symmetric Gaussian probability density to be (see \citep{2006_Gallager_Circular}),
\begin{align}
\mathcal{N}(\bf {x}_i; \boldsymbol{ \Sigma}_{i}) \overset{\Delta}{=}&\frac{1}{{{\pi^{m} \det(\boldsymbol{\Sigma}_{i})}}}e^{-\bf{x}_i^H \boldsymbol{ \Sigma}_{i}^{-1} \bf{x}_i} \label{eqn:g1},\\
\intertext{where $H$ denotes the  conjugate transpose operator and $\boldsymbol{ \Sigma}_{i}$ denotes the covariance matrix which is defined as}\boldsymbol{ \Sigma}_{i} \overset{\Delta}{=}&E\left[ {\bf {x}_i}{\bf {x}_i}^H\right].
\end{align}
\subsection{EM Algorithm for Complex Domain Circular-Symmetric MoG}
\label{sec:EM}
Hereafter we use the term `circular-symmetric MoG' which is to be understood as a mixture of densities with Gaussian components having circular symmetry property.
We model the patches of the phase image using a circular symmetric MoG, i.e.,
\begin{equation}	
p_{\bf X}(\bf x_i)=\sum_{k=1}^{K}\alpha_k \mathcal{N}(\bf {x}_i; \boldsymbol{ \Sigma}_{k}), \label{eqn:MoG}
\end{equation}
where $K$ is the number of components, and $\alpha_k$ and $ \boldsymbol{\Sigma}_k $ are the mixing coefficient and the sample covariance matrix of the $k^{\text{th}}$ component of the MoG respectively. The EM algorithm shown below is used to learn these parameters from the noisy patches.  ${\bf z}_i = {\bf x}_i + {\bf n}_i$.  Assuming that the noise is also circular symmetric Gaussian with covariance matrix $\sigma^2\bf I$, then we have $ p_{Z}(\bf z_i)=\sum_{k=1}^{K}\alpha_k \mathcal{N}(\bf {z}_i; \boldmath{\Gamma}_k )$, where $\boldmath{\Gamma}_k =\boldmath{\Sigma}_k+\sigma^2{\bf I} $.\\
\noindent\rule{13cm}{1.0pt}\\
\textbf{The EM Algorithm}\\
\noindent\rule{13cm}{1.0pt}\\ \\
\line(0,4){190}
\vspace{-200pt}

\begin{enumerate}
\setlength\itemsep{0.1mm}
\item \textbf{Initialization: } Initialize the parameters $\widehat{\alpha} =\lbrace \widehat{\alpha}_k \rbrace_{k=1}^K$ and $\widehat {\boldmath{\Gamma}}= \lbrace \widehat {\boldmath{\Gamma}}_k \rbrace_{k=1}^K$  and evaluate the initial value of the log likelihood.
\item \textbf{E-STEP: } Evaluate the posterior probabilities using the current parameter values\\
 $\gamma_{ik}:=\frac{\widehat{\alpha}_k \mathcal{N}(\bf {z}_i; \widehat {\boldmath{\Gamma}}_k)}{\sum_{j=1}^{K} \widehat{\alpha}_j \mathcal{N}(\bf {z}_i; \widehat {\boldmath{\Gamma}}_j)} $, \ for $i$=1, \dots, $N$   and $k$=1, \dots, $K$\\
 $N_k:=\sum_{i=1}^{N_p}\gamma_{ik}$
\item \textbf{M-STEP: }Re-estimate the parameters using the current posterior probabilities
\begin{enumerate}
\item $\widehat {\boldmath{\Gamma}}_k:=\frac{1}{N_k}\sum_{i=1}^{N_p}\gamma_{ik}\bf{z}_{i}\bf{z}_{i}^H $,\ \ \  for $k=1,2...K$ 
\item $\widehat{\alpha}_k:=\frac{N_k}{N_p}$,\ \ \  for $k=1,2...K$
\end{enumerate}
\item \textbf{Log Likelihood Evaluation:}\\
$ \text{ln} \cbr{ p_{Z}(\bf {z})} :=\sum_{i=1}^{N_p} \text{ln} \sum_{k=1}^{K} \widehat{\alpha}_k \mathcal{N}(\bf {z}_i; \widehat {\boldmath{\Gamma}}_k)  $
\item \textbf{Convergence Check:} Check for convergence of log likelihood. If
the convergence criterion is not satisfied return to step 2, else stop the EM algorithm.
\end{enumerate}
Since $\widehat {\boldmath{\Gamma}}_k \simeq \widehat {\boldsymbol{\Sigma}}_{k}+\sigma^2 \bf{I}, \text{for} \  k=1,...K, $ 
$\widehat {\boldsymbol{ \Sigma}}_{k}$ is estimated as $\widehat {\boldsymbol{ \Sigma}}_{k}= \widehat {\boldmath{\Gamma}}_k-\sigma^2 \bf {I}$.
We use eigen value decomposition and a simple threshold function to get positive definite $\widehat {\boldsymbol{\Sigma}}_k$, i.e., 
$\widehat {\boldsymbol{ \Sigma}}_{k}=\bf{U}\nbr{{\bf {S}- \boldsymbol{\sigma}^{2}}\bf{I}}_{+}\bf{U}^H$, where the matrices $\bf{U}$ and $\bf{S}$ are eigenvector and eigenvalue (diagonal) matrices of $\widehat {\boldmath{\Gamma}}_k$ and $x_+:=max \nbr{0,x}$.


\subsection{MMSE Estimation of the Clean Patches}
\label{sec:mmse}
\begin{flalign}
\intertext{The posterior probability of a patch ${\bf x}_i$ given ${\bf z}_i$, for $i=1, \dots, N_p$, is given by}
p_{X|Z}(\p{x}|\p{z})&=\frac{p_{Z|X}(\p{z}|\p{x})p_X(\p{x})}{p_Z(\p{z})}=\frac{p_{Z|X}(\p{z}|\p{x})\sum_{k=1}^{K}\alpha_k p_X^k(\p{x})}{\sum_{k=1}^{K}\alpha_k p_Z^k(\p{z})} \\
p_X^k(\p{x})&:=\mathcal{N}(\bf {x}_i;  \widehat{\boldsymbol{\Sigma}}_{k}), \\
p_Z^k(\p{z})&:=\mathcal{N}(\bf {z}_i; \widehat{\boldsymbol{\Sigma}}_{k}+\sigma^2\bf{I})
\intertext{The MMSE Estimate (Posterior Mean) of a patch $\p{x}$ is given by}
\widehat{\bf{x}}_{i,\text{mmse}}&=\int_{-\infty}^{\infty} \p{x} p_{X|Z}(\p x| \p z) d \p{x} =\int_{-\infty}^{\infty} \p{x} \frac{p_{Z|X}(\p{z}|\p{x})\sum_{k=1}^{K}\alpha_kp_X^k(\p{x})}{\sum_{k=1}^{K}\alpha_kp_Z^k(\p{z})} d \p{x}, \label{eqn:mmsepere}
\intertext{which on further calculation gives,}
\widehat{\bf{x}}_{i,\text{mmse}}&=\sum_{k=1}^{K} \widehat{\alpha}_k \widehat{\bf{x}}^k_{i,\text{mmse}},\\
\widehat{\alpha}_k  &=  \frac{\alpha_k p_Z^k(\p{z})}{P_Z(\p{z})}, \ \ \widehat{\bf{x}}^k_{i,\text{mmse}}=\widehat{\boldsymbol{\Sigma}}_{k}(\widehat{\boldsymbol{\Sigma}}_{k}+\sigma^2 \bf I)^{-1} \p{z}.
\end{flalign}

For more details of the derivation of the posterior mean for the MoG component $k$, i.e. $\widehat{\bf{x}}^k_{i,\text{mmse}}$, refer \citep{1993_Kay_Fundamentals}
\section{NL-Averaging}
After the MMSE denoising step, we obtain a set of filtered patches in which the noise is largely attenuated. However, there is still room to reduce the noise by further exploiting the image self-similarity using non-local (NL)-averaging in the spirit of \cite{2005_Buades_image}. 
 For each $\widehat{\bf{x}}_{i,\text{mmse}} \in \mathbb{C}^m$, the NL-average is found as
\begin{align}
\widehat{\bf{x}}_{i}=\sum_{j=1}^{N_p}\widehat{\bf{x}}_{j,\text{mmse}}e^{-\frac{\norm{\widehat{\bf{x}}_{i,\text{mmse}}-\widehat{\bf{x}}_{j,\text{mmse}}}_F^2
}{h^2}} \ \text{for} \  i=1,2,..., N_p, \label{NL1}
\end{align}
where $h$ is a parameter tuned experimentally ($h=0.48\sigma$) and $\norm{.}_F^2$ is the Frobenius norm. As this is computationally expensive, we restrict the averaging process within a fixed size neighbourhood (square window of size $11 \times 11$) of each patch. We thus compute  
\begin{align}
\widehat{\bf{x}}_{i}=\sum_{\widehat{\bf{x}}_{j,\text{mmse}} \in NB_i}\widehat{\bf{x}}_{j,\text{mmse}}e^{-\frac{\norm{\widehat{\bf{x}}_{i,\text{mmse}}-\widehat{\bf{x}}_{j,\text{mmse}}}_F^2
}{h^2}} \ \text{for} \  i=1,2,..., N_p, \label{NL2}
\end{align}
where $NB_i$ denotes the set of patches which are near to the patch $i$ in their respective positions in the original image. We remark that compared with the original NL-filtering scheme introduced in \cite{2005_Buades_image}, we use  denoised patches to compute the Euclidean distances, which has a positive impact in the final results.

\section{Experiments and Results}
\label{sec:exp}
   
The performance of the proposed algorithm is compared with the state-of-the-art techniques for Inphase Image denoising, namely SpInphase \cite{2015_Hongxing_Interferometric} and WFT \cite{2007_Kemao_Twodimensional}. As our interest is on unsupervised or mildly supervised scenarios, we do not consider any deep learning methods for the quality comparison. We adopt a quality measure termed as \textit{peak signal-to-noise ratio} (PSNR) to compare the performance of different algorithms \cite{2015_Hongxing_Interferometric}.
\vspace{-0.3cm}
\begin{equation}
PSNR:=10\log_{10}\frac{4N\pi^2}{ \norm{\mathcal{W}(\widehat{\boldsymbol{\upphi}}_{2\pi}-\boldsymbol{\upphi}_{})}_F^2} \ \  [\text{dB}],
\end{equation}
where $\boldsymbol{\upphi}_{}$ is the true phase (unwrapped), $\widehat{\boldsymbol{\upphi}}_{2\pi}$ is the estimated wrapped phase and $\mathcal{W}$ is the wrapping operator defined in \eqref{eqn:wrap}. As already mentioned in Section \ref{sec:intro}, the process of interferometric phase estimation is usually accomplished by Phase Denosing and Phase Unwrapping. The success of Phase Unwrapping depends crucially on the quality of the denoised interferometric phase. To account this factor, we unwrap the denoised InPhase image using the state-of-the-art PUMA algorithm \cite{2007_Bioucas_Phase}. To measure the quality of the unwrapped denoised phase ($\widehat{\boldsymbol{\upphi}}$), as in \cite{2015_Hongxing_Interferometric}, we define a set of image pixels having error less than $\pi$ compared to the true phase image ($\boldsymbol{\upphi}$), i.e., $I:=\cbr{i:|\widehat{\upphi}_i - \upphi_i|\leq \pi, i=1,...N}$. Based on this set, we define \textit{number of error larger than $\pi$} (NELP) and a new \textit{peak signal-to-noise ratio} ($\text{PSNR}_a$) as
\vspace{-0.3cm}
\begin{align}
\text{NELP} :=N-|I|;  \ \ 
PSNR_a :=10\log_{10}\frac{4N\pi^2}{ \norm{(\widehat{{\upphi}}_{I}-{\upphi}_{I})}_F^2} \ \  [\text{dB}],
\end{align}
where the notation ${\upphi}_{I}$ stands for the restriction of ${\upphi}$ to the set $I$ \cite{2015_Hongxing_Interferometric}. The number of components of the MoG in all the following experiments are selected heuristically for optimal performance. Also, the patches considered are of the size $10\times10$ (i.e., $m=100$). Although the time cost of the algorithm is not considered in the performance evaluation, we would like to mention that the computational complexity of the proposed algorithm (MMSE and NL-Averaging) is linear with the number of pixels considered. 
\vspace{-0.2cm}
\subsection{Experiments Conducted on Simulated Data Set}
\begin{figure}[H]
\begin{center}
\includegraphics[height=2.5cm,width=12cm]{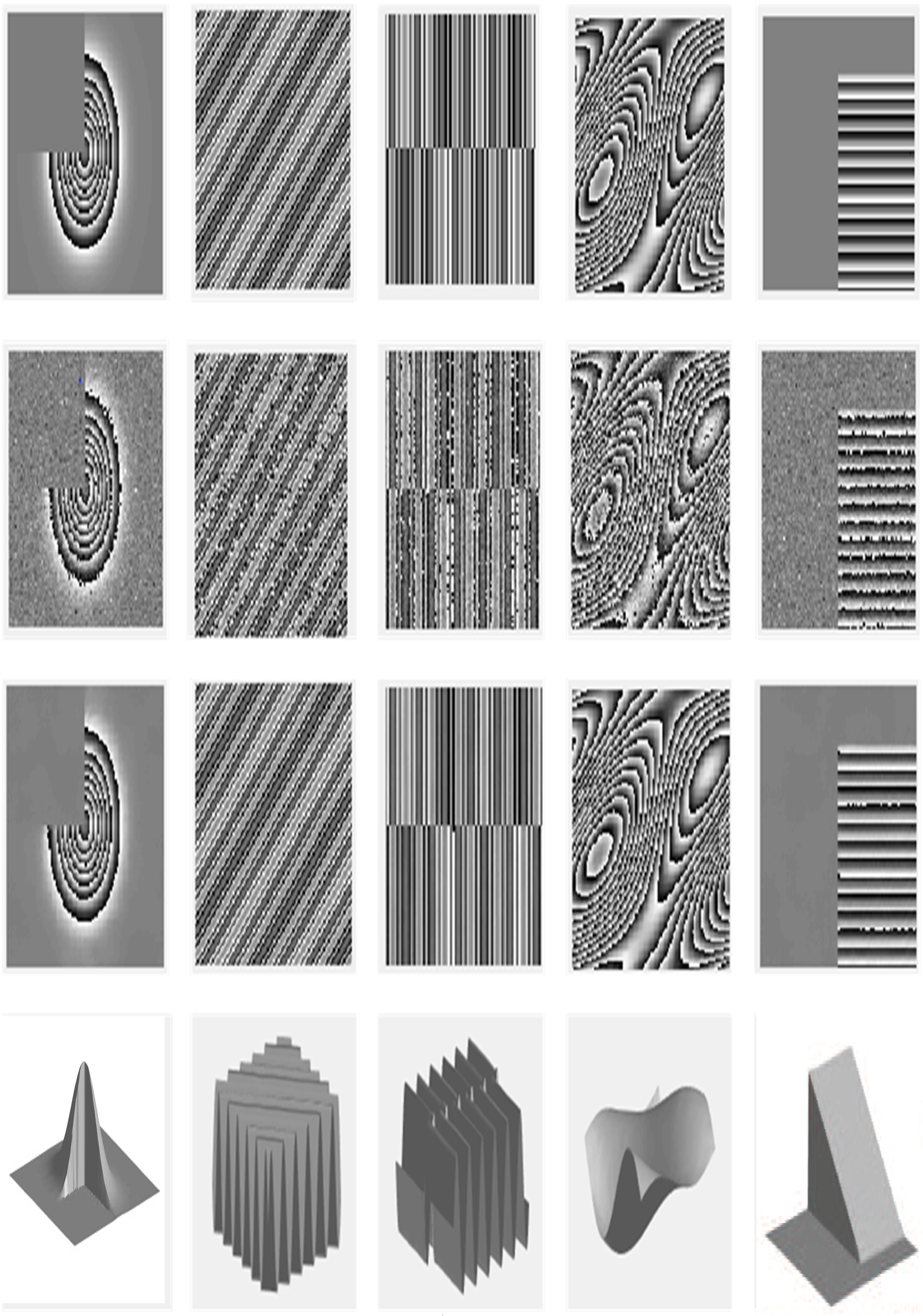}
\end{center}
\caption{\scriptsize From top to bottom:  the true interferometric phase, the noisy interferometric
phase for $\sigma σ = 0.5$, the InPHASE estimate, unwrapped surface. From the left to the right: Truncated Gaussian, Sinusoidal, Discontinuous
Sinusoidal, Mountains, Shear Planes.}
\label{puma}
\end{figure}

Five different data sets are created to model five different topologies. The size of the data sets, shown in Fig. \ref{surface}, is $100 \times 100$. The observed data is generated according to \eqref{eqn:1}. The interferometric phases shown in the subfigures are represented in gray level: black represents $-\pi$ and white represents $\pi$. Each of the above surfaces is  considered for the phase image $\boldsymbol \upphi$. Here we focus on the inference of interferometric phase and the detailed statistical characterisation of the amplitude image is beyond the scope of this paper. But we remark that the propsed approach gives competitive results even for a constant amplitude signal model, though the MoG is not the best density model for such cases. In the following experiments conducted on the simulated data sets, we consider a smoothly varying non-negative surface, i.e., the mountain surface of Fig. \ref{surface} (d), as the amplitude image ($\bf a $). We present two sets of experiments with these simulated data set. In the first set of experiments, which is pre-learning experiment (pl),  the learning (MoG: 30 components, SpInphase: 512 atoms ) is done from the five clean images combined together. The second set of experiments are self-learning (sl) experiments in which the learning (MoG: 15 components, SpInphase: 256 atoms ) is done from the noisy data itself. Also to test the algorithm, noisy versions of each of the aforementioned surfaces are considered with $\sigma \in \lbrace0.3, 0.5, 0.7, 0.9, 1 \rbrace$. Table \ref{table1} shows the performance of the proposed algorithm in comparison with the state-of-the-art. The complete process of InPhase denoising and unwrapping \cite{2007_Bioucas_Phase} are illustrated in Fig. \ref{puma}.

\begin{figure}[h!]
\begin{center}
\begin{tabular}{ccccc}
\includegraphics[height=1.1cm,width=2cm]{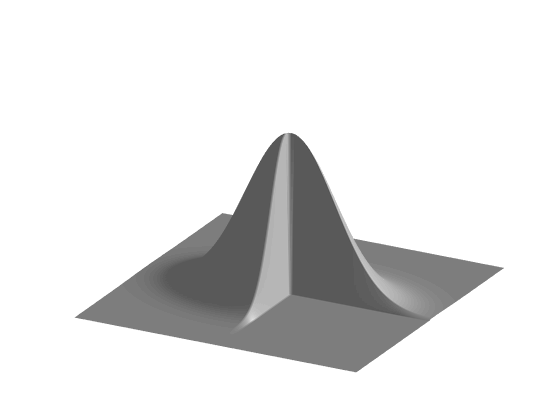} &
\includegraphics[height=1.1cm, width=2cm]{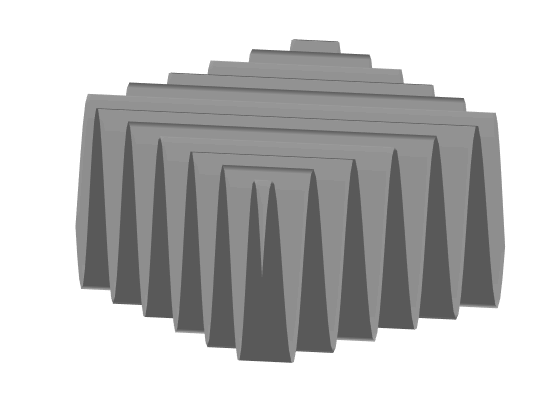} &
\includegraphics[height=1.1cm,width=2cm]{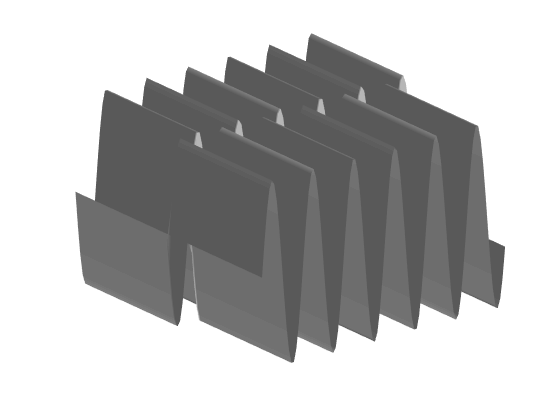} &
\includegraphics[height=1.1cm, width=2cm]{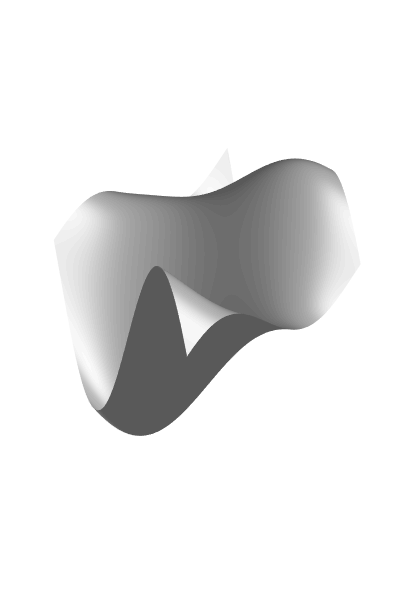} &
\includegraphics[height=1.1cm,width=2cm]{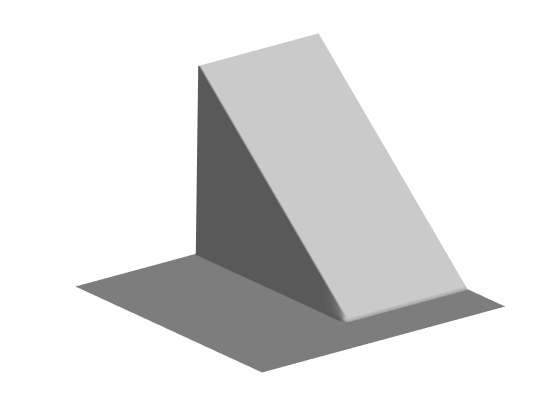}\\
 \scriptsize{(a) Truncated Gaussian} &  \scriptsize{(b) Sinusoidal} & \scriptsize{(c) Discontinuous sinusoidal} &  \scriptsize{(d) Mountainous }&\scriptsize{(e)Shear Planes}
\end{tabular}
\end{center}
\caption{\scriptsize Simulated data sets}
\label{surface}
\end{figure}

\vspace{-0.5cm}
Table \ref{table1} shows that the proposed approach is highly competitive with the state-of-the-art even for the surfaces like discontinuous sinusoidal where the amount of discontinuity is very high. Also, the very low NELP values, in most of the cases, show that the wrapping discontinuities are preserved very well which is a crucial factor that maintains the quality of the unwrapped InPhase images.

\begin{table}[H]
\begin{center}
  \tiny
    \begin{tabular}{r|c|cc|cc|c|cc|cc|c|cc|cc|c}
    \multicolumn{1}{r|}{\multirow{2}[1]{*}{Surf.}} & \multirow{2}[1]{*}{$\sigma$} & \multicolumn{5}{c|}{PSNR (dB)} & \multicolumn{5}{c|}{NELP} & \multicolumn{5}{c}{PSNRa (dB)} \\
    \multicolumn{1}{r|}{} &       & \multicolumn{1}{c}{MoG} & \multicolumn{1}{c|}{SP} & \multicolumn{1}{c}{MoG} & \multicolumn{1}{c|}{SP} & WFT 
    			          & \multicolumn{1}{c}{MoG} & \multicolumn{1}{c|}{SP} & \multicolumn{1}{c}{MoG} & \multicolumn{1}{c|}{SP} & WFT   
    				  & \multicolumn{1}{c}{MoG} & \multicolumn{1}{c|}{SP} & \multicolumn{1}{c}{MoG} & \multicolumn{1}{c|}{SP} & WFT\\
    				  
    \multicolumn{1}{r|}{} &       &  \multicolumn{2}{c|}{(pl)} & \multicolumn{2}{c|}{(sl)}   & {} 
    			          &  \multicolumn{2}{c|}{(pl)} & \multicolumn{2}{c|}{(sl)}   & {}   
    				  &  \multicolumn{2}{c|}{(pl)} & \multicolumn{2}{c|}{(sl)}   & {}\\
    \midrule
    \midrule
    \multirow{2}[1]{*}{Trunc. }    & 0.3   & \B{45.31} & 45.08 	& \B{44.51} & 44.43 	& 44.06  	& \B{0}   & \B{0}  & 1      &  \B{0}    & \B{0}    		& \B{45.31} & 45.08 	& \B{44.52}  & 44.43 	& 44.06 \\
               		           & 0.5   & 41.21 & \B{42.16} 	& 39.81 & \B{42.38} 	& 40.24  	& 11      & \B{0}  & \B{0}  &  \B{0}    & \B{0}    		& \B{42.22} & 42.16 	& 39.81  & \B{42.38} 	& 40.24 \\
    \multirow{2}[1]{*}{Gauss.}     & 0.7   & 38.14 & \B{40.75} 	& 36.55 & \B{39.00} 	& 37.40  	& 9       & \B{0}  & 6      &  \B{0}    & \B{0}   		& 40.10 & \B{40.75}	& 36.92  & \B{39.00} 	& 37.40 \\
			           & 0.9   & 37.77 & \B{39.03} 	& 36.28 & \B{38.14} 	& 36.58  	& 43      & \B{0}  & 36     &  \B{0}    & \B{0}                 & \B{39.72} & 39.03 	& 37.83  & \B{38.14} 	& 36.58 \\
    \midrule
    \midrule
    \multirow{4}[0]{*}{Sinu.} 
				   & 0.3   & 47.76 & \B{48.69} 	& \B{51.72} & 47.80 	& 40.45  	& \B{0} & \B{0} & \B{0}  & \B{0} & \B{0} 			& 47.76 & \B{48.69} 	& \B{51.72}	 & 47.80	& 40.45 \\
				   & 0.5   & 46.43 & \B{48.06} 	& \B{46.86} & 42.84 	& 35.95  	& \B{0} & \B{0} & \B{0}  & \B{0} & \B{0} 			& 46.43 & \B{48.06} 	& \B{46.86}	 & 42.84	& 35.95 \\
				   & 0.7   & 41.88 & \B{44.12} 	& \B{43.38} & 37.57 	& 33.15  	& \B{0} & \B{0} & \B{0}  & \B{0} & \B{0} 			& 41.88 & \B{44.12} 	& \B{43.38}	 & 37.57	& 33.15 \\
				   & 0.9   & 40.61 & \B{42.87} 	& \B{40.78} & 35.99 	& 29.78  	& \B{0} & \B{0} & \B{0}  & \B{0} & \B{0} 			& 40.61 & \B{42.87} 	& \B{40.78}	 & 35.99	& 29.78 \\
    \midrule
    \midrule     
    \multirow{2}[0]{*}{Sinu.}      & 0.3   & \B{45.05} & 43.65 	& 45.47 & \B{46.18} 	& 39.11  	& \B{0} & \B{0} & \B{0} & \B{0} & \B{0}  			& \B{45.05} & 43.65 	& 45.47	 & \B{46.18}	& 39.11 \\
			           & 0.5   & \B{42.35} & 40.60 	& 42.68 & \B{43.00} 	& 35.38  	& \B{0} & \B{0} & \B{0} & \B{0} & \B{0}  			& \B{42.35} & 40.60 	& 42.68	 & \B{43.00}	& 35.38 \\
 \multirow{2}[0]{*}{Disc.}         & 0.7   & \B{40.19} & 39.41 	& 39.75 & \B{41.92} 	& 33.04  	& \B{0} & \B{0} & \B{0} & \B{0} & 1  			        & \B{40.19} & 39.41 	& 39.75	 & \B{41.92}	& 33.16 \\
                                   & 0.9   & \B{37.36} & 37.27 	& \B{39.21} & 37.54 	& 30.55  	& \B{0} & \B{0} & \B{0} & \B{0} & 11  		                & \B{37.36} & 37.27 	& \B{39.21}	 & 37.54	& 30.80 \\
     \midrule
    \midrule
    \multirow{4}[0]{*}{Mount.} 
				   & 0.3   & 43.05 & 42.71 	& 40.98 & 43.06 	& \B{44.60} 	& \B{0} & \B{0} & \B{0} & \B{0} & \B{0}  			& 43.05 & 42.71 	& 40.98	 & 43.06	& \B{44.60} \\
				   & 0.5   & 40.22 & 40.16 	& 38.23 & 40.43 	& \B{41.47} 	& \B{0} & \B{0} & \B{0} & \B{0} & \B{0}  			& 40.22 & 40.16 	& 38.23	 & 40.43	& \B{41.47} \\
				   & 0.7   & 38.57 & 38.50 	& 36.55 & 38.72 	& \B{39.20} 	& \B{0} & \B{0} & \B{0} & \B{0} & 1  			        & 38.57 & 38.50 	& 36.55	 & 38.72	& \B{39.20} \\
				   & 0.9   & 36.28 & 37.51 	& 35.08 & 36.35 	& \B{37.85} 	& \B{0} & \B{0} & \B{0} & \B{0} & \B{0}  			& 36.28 & 37.51 	& 35.08	 & 36.35	& \B{37.85} \\
    \midrule
    \midrule
    \multirow{2}[1]{*}{Shear}      & 0.3   & \B{53.39} & 49.50 	&\B{52.45} & 52.37   	& 45.75    	& \B{0} & \B{0} & \B{0}  & \B{0} & \B{0}  		& \B{53.39} & 49.50 	& \B{52.45}  & 52.37 	& 45.75 \\
                                   & 0.5   & \B{49.23} & 47.35 	&\B{49.24} & 47.23 	        & 41.86    	& \B{0} & \B{0} & \B{0}  & \B{0} & \B{0}  		& \B{49.23} & 47.35 	& \B{49.24}  & 47.23 	& 41.86 \\
    \multirow{2}[1]{*}{Plane}      & 0.7   & \B{46.18} & 45.27 	&\B{45.41} & 42.86   	& 39.31    	& \B{0} & \B{0} & \B{0}  & \B{0} & \B{0}  		& \B{46.18} & 45.27 	& \B{45.41}  & 42.86 	& 39.31 \\
                                   & 0.9   &   42.87 & \B{43.81} &\B{43.43} & 41.46  	& 37.22    	& \B{0} & \B{0} & \B{0}  & \B{0} & \B{0}  		&   42.87 & \B{43.81} 	& \B{43.43}  & 41.46 	& 37.22 \\
    \bottomrule
    \bottomrule
    \end{tabular}%
\end{center}
\caption{ \scriptsize Performance Indicators for the surfaces shown in Fig.\ref{surface}. pl: pre-learned, sl: self-learned}
\label{table1}
\end{table}%
\vspace{-0.5cm}

\subsection{Experiments Conducted on Real MRI Data}

\vspace{-0.4cm}

\begin{figure}[H]
\begin{center}
\includegraphics[height=2cm,width=12cm]{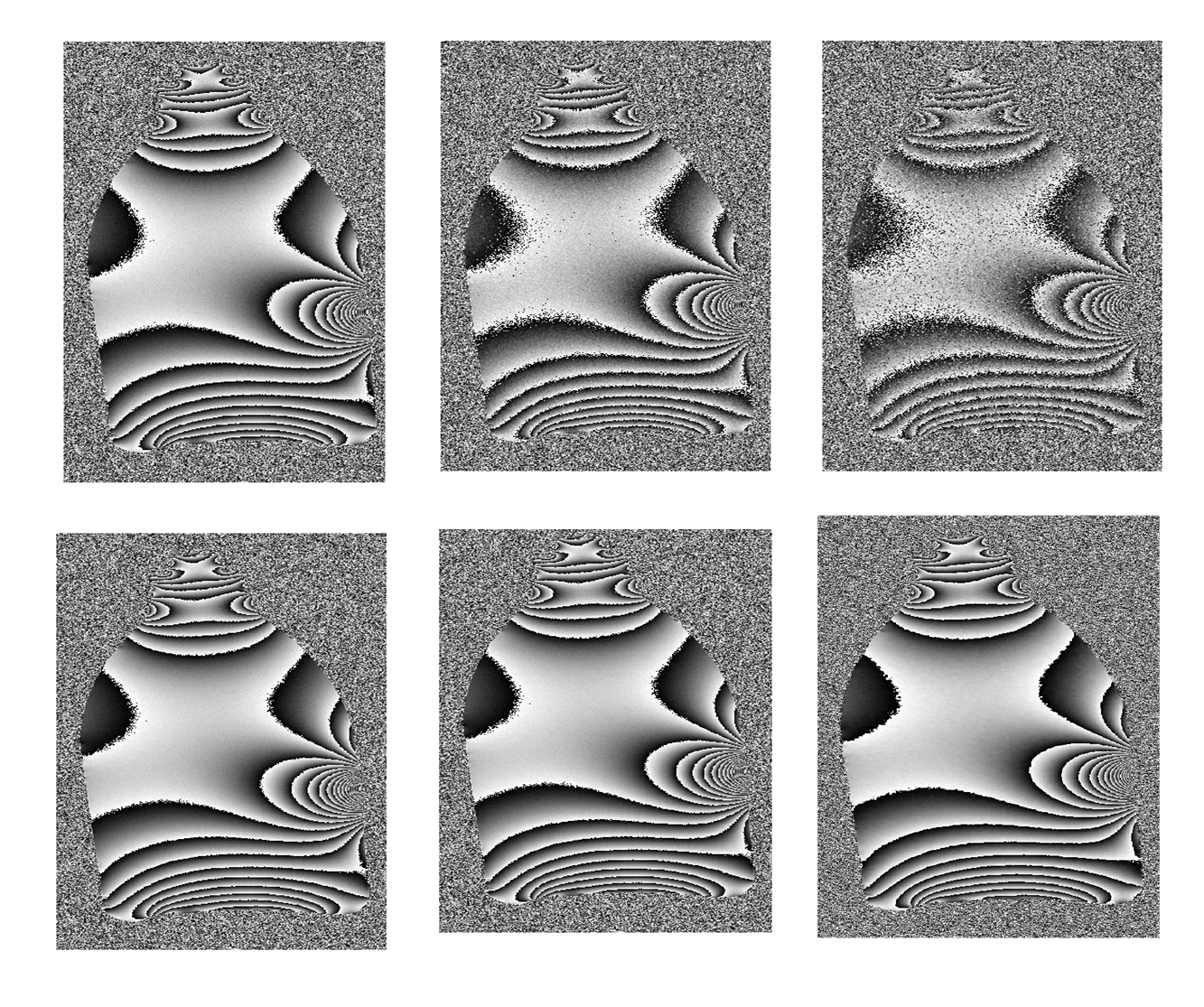}
\end{center}
\caption{\tiny Top: MRI InPhase images of a bottle with increasing noise level from left to right. Bottom:
Corresponding MoGInPhase estimates.}
\label{mri}
\end{figure}

Fig. \ref{mri} shows MRI interferometric phase images of a bottle filled with liquid. The MRI images, of size 512 $\times$228,
were acquired using decreasing scan times from the left to right. A 20-component MoG is used  and the parameters are learned from the noisy data. It is assumed that the complex valued signal varies smoothly and the noise is estimated from the first order horizontal and vertical differences. The noise variance estimated are 0.158, 0.382, 0.720.  The results for the proposed algorithm in comparison with the state-of-the-art are tabulated in Table \ref{table_mri}. The Table \ref{table_mri} shows that the proposed method is well suited for the real MRI InPhase data and is highly competing with the state of the art in terms of $PSNR$, $NELP$ and $PSNR_a$   

\begin{table}[htbp]
  \begin{center}
    \tiny
    \begin{tabular}{c|ccc|ccc|ccc}
    \multicolumn{1}{c|}{\multirow{2}[1]{*}{sigma}} & \multicolumn{3}{c|}{PSNR (dB)} & \multicolumn{3}{c|}{NELP} & \multicolumn{3}{c}{PSNRa (dB)} \\
    \multicolumn{1}{c|}{} & \multicolumn{1}{c}{MoG} & \multicolumn{1}{c}{SPIn
Phase} & WFT   & \multicolumn{1}{c}{MoG} & \multicolumn{1}{c}{SPIn
Phase} & WFT   & \multicolumn{1}{c}{MoG} & \multicolumn{1}{c}{SP
InPhase} & WFT \\
    \midrule
    \midrule
    0.153 & \B{35.900} & 35.487   & 35.777   & 164   & \B{141}   & 417     & \B{36.120}   & 35.489   & 35.781   \\
    0.382 & \B{34.842} & 34.830   & 34.601   & \B{176}   & 448   & 499     & \B{34.899}   & 34.847   & 34.610   \\
    0.72  & 32.371 & 32.520   & \B{32.684}   & 447   & 274   & \B{143}     & \B{33.000}   & 32.606   & 32.689   \\
    \bottomrule
    \bottomrule
    \end{tabular}%
    \end{center}

\caption{ \scriptsize Performance Indicators for the MRI images shown in Fig.\ref{mri}. MoG (20 components) is learned from the noisy images}
\label{table_mri}
\end{table}%
\vspace{-0.5cm}
\subsection{Experiments Conducted on InSAR Data}
Experiments are conducted using the InSAR data distributed with the book \citep{1998_Ghiglia_Two}. The data sets were generated based on a real digital elevation model of mountainous terrain around Longs Peak and Isolation Peak, Colorado, using a high-fidelity InSAR simulator. The detailed description of the simulator are given in \cite[Chapter~3]{1998_Ghiglia_Two} and we conduct the same set of experiments as described in \cite{2015_Hongxing_Interferometric}. The $PSNR$ values (in dB) obtained for MoGInPhase, SPInPhase and WFT are respectively (24.06, 26.02, 20.42) for the Longs Peak and (23.40, 24.13, 20.77) for the Isolation Peak.  In these two experiments, the proposed algorithm performs very close to the SPInPhase, although a bit below. We are not providing the $NELP$ values for this experiment as the unwrapping of InSAR data with quality map is beyond the scope of this work. 

\begin{figure}[H]
\begin{center}
\includegraphics[height=3cm,width=12cm]{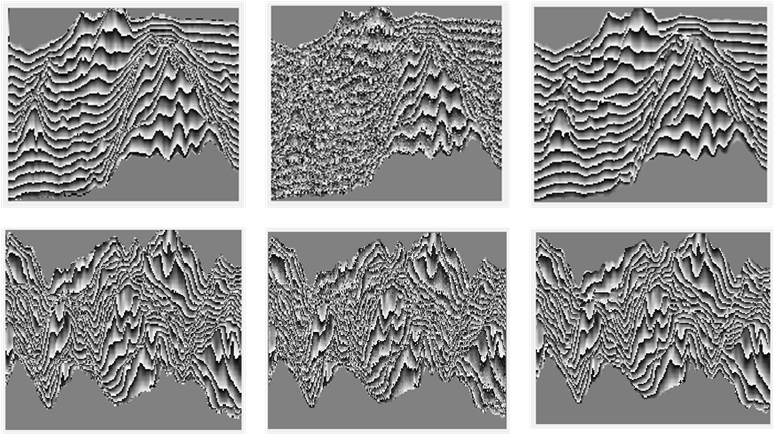}
\end{center}
\caption{\tiny Estimation results for the real digital elevation model. Top: Longs Peak, Bottom: Isolation Peak, Left to Right : Original Wrapped InPhase image, InPhase image with InSAR noise, MoGInPhase estimate in the order}
\label{insar}
\end{figure}

\section{Conclusion}
\label{sec:concl}
This paper introduced an effective two-stage algorithm for interferometric phase image denoising. The two stages, i.e., MoGInPhase and NL-averaging are designed to exploit the non-local self-similarity of the phase images. The experiments conducted on real and simulated data show results which are competitive with the state-of-the-art techniques. One of the relevant contributions of the proposed algorithm is that it opens the door to the exploitation of ``learned priors" from the specified classes of the interferometric phase images, which can then be used in various phase inverse problems.

\bibliographystyle{IEEEbib}
\bibliography{refs}
\end{document}